\documentclass{article}
\usepackage{spconf,amsmath,graphicx}
\usepackage[hidelinks]{hyperref}
\usepackage{xcolor}

\title{Technology Pipeline for Large Scale Cross-Lingual Dubbing\\of Lecture Videos into Multiple Indian Languages}
\makeatletter
\def\@name{{\it Anusha P$^1$, Arun Kumar$^2$, Ashish Seth$^2$, Bhagyashree M$^1$, Ishika Gupta$^1$,} \\{\it Jom Kuriakose$^1$, Jordan Fernandes$^2$, K V Vikram$^1$, Mano R Kumar M$^1$,}\\ {\it Metilda S M$^2$, Mohammad Wajahat$^2$, Mohana N$^1$, Mudit Batra$^2$, Navina K$^1$, Nihal G$^1$, Nithya Ravi$^2$,}\\{\it Pruthwik Mishra$^3$, Sudhanshu S$^1$, Vasista L$^2$, Vandan Mujahida$^3$, Vineeth K$^1$, Vrundha S$^2$ }\\{\it Dipti Mishra$^3$, Hema A Murthy$^1$, Pushpak Bhattacharya$^4$, Srinivasan Umesh$^2$, Rajeev Sangal$^3$}\thanks{Thanks to MeiTY Govt of India for the project: 11(1)/2022-HCC(TDIL)}}
\makeatother
\address{$^1$ SMT Lab, Department of Computer Science and Engineering, IIT Madras, India\\
$^2$ Speech Lab, Department of Electrical Engineering, IIT Madras, India\\
$^3$ IIIT Hyderabad, India \hspace{1cm} $^4$ IIT Bombay, India}

\begin{document}
%
\maketitle
\begin{abstract}
Cross-lingual dubbing of lecture videos requires the transcription of the original audio, correction and removal of disfluencies, domain term discovery, text-to-text translation into the target language, chunking of text using target language rhythm, text-to-speech synthesis followed by isochronous lipsyncing to the original video. This task becomes challenging when the source and target languages belong to different language families, resulting in differences in generated audio duration. This is further compounded by the original speaker's rhythm, especially for extempore speech. This paper describes the challenges in regenerating English lecture videos in Indian languages semi-automatically. A prototype is developed for dubbing lectures into 9 Indian languages. A mean-opinion-score (MOS) is obtained for two languages, Hindi and Tamil, on two different courses. The output video is compared with the original video in terms of MOS (1-5) and lip synchronisation with scores of 4.09 and 3.74, respectively. The human effort also reduces by 75$\%$.
\end{abstract}
\section{Introduction}
\label{sec:intro}

India has a rich linguistic diversity, with 22 official languages belonging to different language families. The medium of higher education, however, is predominantly English. There is a dearth of educational resources in regional languages, resulting in the marginalisation of a large section of society. Recent efforts have attempted to make educational resources available in regional languages using available speech and machine translation technologies to bridge this gap (\url{nptel.ac.in}). While this effort is useful, the audio-visual experience of watching the video in the language of the listener is absent.  

\begin{figure*}
    \centering
    \includegraphics[width=14cm, height=4cm]{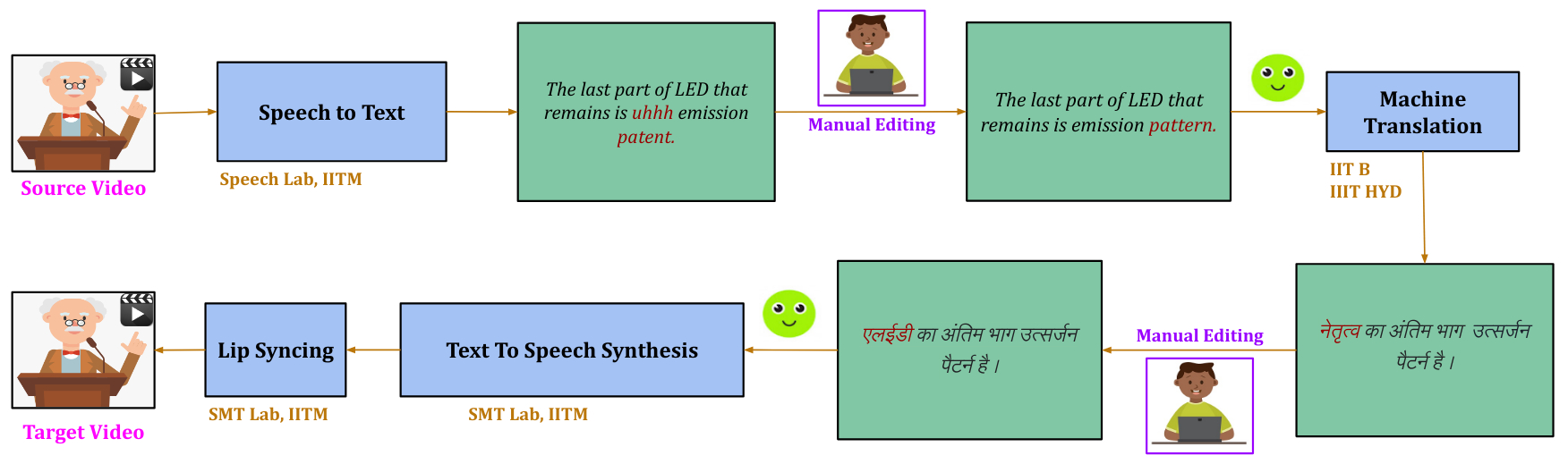}
    \caption{Video to video pipeline for dubbing of technical lectures}
    \label{fig:transcreation}
\end{figure*}

A cross-lingual dubbing system typically involves four modules -- (1) An automatic speech recogniser (ASR) to transcribe the English lectures. (2) A machine translation (MT) system translates the transcribed text to a target language. (3) A text-to-speech (TTS) system synthesises the translated text. (4) A lip-syncing module to synchronise the TTS audio with the original video. The modules involved in dubbing lecture videos are shown in Figure \ref{fig:transcreation}. Each module is prone to errors. Each of the modules leads to various challenges. We now briefly list the challenges and later discuss novel ways in which they are addressed.  

\section{Challenges in transcreation}
\label{sec:challenges}
\subsection{Automatic Speech Recognition (ASR) of lecture videos}
State-of-the-art ASR systems work well for general Indian English but fail for technical lectures. Technical lectures are more or less conversational and are replete with esoteric terms. Moreover, the sentences are often incomplete or ill-formed. This is compounded by the fact that Indians are bilingual and tend to switch between English and their native tongue. The ASR outputs are often time-stamped using voice-activity detection (VAD) to produce SRTs (sub-rip-text), which are used for subtitling the videos.   The SRTs are seldom grammatically correct, which makes the task of translation difficult.
\subsection{What to translate?  What to transliterate?}
While most speech recognition systems are able to transcribe verbatim the content spoken by the speaker, the text is quite unreadable owing to the presence of disfluencies. ASR outputs also often lack proper punctuation. ASR outputs are post-processed to remove disfluencies and summarised to make meaningful sentences for machine translation.   Many of the technical words in these lectures need to be transliterated rather than translated, owing to the nonavailability of these terms in the vernacular, or the terms are rather esoteric.   This requires the identification of domain terms in the source. Occasionally, even non-domain terms must be kept as is. For example, consider an example from the programming language C:
\begin{verbatim}
        for (i = 0; i< n-1; i++)
            a++;
\end{verbatim}
Here the variable {\bf i} could be mistaken by the translator to mean a person referring to himself/herself. Ideally, the entire {\bf for} loop should not be translated.   

\subsection{Punctuation and Indian languages}
Indian languages seldom include punctuation. Unlike English, the meaning of an Indian language sentence does not change with punctuation or word order. Indian languages are primarily phrase-based and are made up of very long sentences. We need prosodic markers to ensure that the language rhythm is captured for the text-to-speech synthesiser in the target language.  
\subsection{Isochronous lip syncing}
Word order differences between English and Indian languages, and the length of translated text lead to a mismatch in the duration of the original audio and the synthesised audio. The audio and video need to be synced back so that the audio-visual experience of the lecture is preserved even in the target language.  

In the following Sections, we develop novel methods to address each of the aforementioned issues.
Section \ref{sec:asr} discusses methods to alleviate the issues in the transcription of technical lectures.
Section \ref{sec:dt} discusses the issue of domain term detection, and Section \ref{sec:mt} discusses words and phrases that should be transliterated rather than translated. These two steps are crucial. Further, most technical lectures in Indian languages are always bilingual. It is common for faculty to switch between English and their native tongue. English is a compulsory language in elementary school, the reason why it is common that most technical terms are absorbed into the native language vocabulary. Section \ref{sec:punctuation} studies the speaker specific and language specific rules that need to be honoured so that the prosody of the source speaker and target language are preserved. Section \ref{sec:lipsync} discusses methods to address the issue of ensuring that lip-syncing is more or less accurate. language specific methods are used here. Section \ref{sec:expts} discusses our experimental methodology. All experiments were performed on NPTEL (\url{nptel.ac.in}) video lectures. In Section \ref{sec:conclusion} we report our efforts, discuss the limitations of the work, and the suggested course of future efforts.

\section{Related Work}
\label{sec:related}
Earlier efforts in lip-syncing focused on different techniques for aligning source video and target audio or text. 
In recent work by \cite{federico-etal-2020-speech, Federico2020, virkar22_interspeech}, TED talks are automatically dubbed using a prosodic alignment module. The attention mechanism used in \cite{oktem19_interspeech} tries to find a plausible phrasing for the translated text.
Changing the speed of synthetic speech to fit into the subtitle duration is attempted in \cite{Matuosek2012,Howto3629421:online}.
An end-to-end audio-visual translation system trained on thousands of hours of data from all domains  \cite{DBLP:journals/corr/abs-2011-03530} and adapted to a specific domain and speaker is reported in \cite{DBLP:journals/corr/abs-2011-03530}. Wav2Lip \cite{jawahar2019} produces good lip-syncing provided that a face is present in the entire video. In \cite{xie2021}, few-shot dubbing is attempted, where arbitrary audio is synced to a talking head. An editor for aiding the lip-syncing and multilingual dubbing is developed in \cite{10.1145/3490035.3490284}. In \cite{https://doi.org/10.48550/arxiv.2206.04523}, the target audio recreates the original emphases mapped from the original sentence along with voice conversion to source speaker's voice.
Nevertheless, no attempt has been made to address the issues in dubbing technical lectures into different languages, especially languages that do not share a script.

\section{Automatic speech recognition}
\label{sec:asr}
In this work, NPTEL (\url{nptel.ac.in}) lecture videos for which manual transcriptions are available are used to train the speech recognition systems from scratch. Videos from different domains, including computer science, electrical engineering, mechanical engineering, basic sciences and humanities, are chosen for training the end-to-end ASR model. Since we may encounter new technical terms and domains, and occasionally the speakers do delve into the native tongue, the system does not use a language model. A state-of-the-art conformer model is trained with $12$ encoder layers and $6$ decoder layers using 5000 byte-pair encodings (BPEs) as targets.   Since Indian English mannerisms are similar, the English text obtained is more or less verbatim despite significant differences in accents. The word error rate (WER) of the transcription is quite good $<10\%$, which is much better than that of many commercially available paid services which result in about 25-30\% WER.   While the transcription is accurate, the text needs to be post-processed for disfluency removal and chunked appropriately so that the text translation is accurate. The disfluent speech is first converted to fluent speech using text summarization. Manual transcription of an hour-long video takes about two days, while the manual effort required after ASR takes about 4 hours, leading to a significant reduction in turnaround time.

\section{Domain term detection}
\label{sec:dt}
Technical lectures have a large number of mathematical symbols, equations and abbreviations. Handling such symbols without any loss of information is a challenging task in the MT and TTS stages of the V2V pipeline, even if the ASR is able to predict the text/symbols correctly. The domain terms need to be discovered for each technical domain separately for each topic. The term ``Python" can have different meanings in Computer Science and life sciences. Identification of such terms based on relevant domain concepts for a given document is a highly challenging task. These terms usually lead to different vocabulary selections in machine translation. As a special case to the domain term, we also include mathematical symbols, variables and equations as technical expressions for the identification. The automatic system identifies domain terms and their domains by using both unsupervised (i.e TextRank\cite{mihalcea2004textrank} and TFIDF\cite{joachims1996probabilistic}) and  supervised (Token classification using ALBERT\cite{Lan2020ALBERT}) methods built on inhouse domain dictionaries and TermTraction corpora \footnote{\url{https://ssmt.iiit.ac.in/TermTraction}}. An F1-score of  ~60\% was obtained for domain term identification for a finite set of domains. The dictionary is continuously evaluated and augmented.

\section{Machine translation for the technical domain}
\label{sec:mt}
Terminology integrated specialised machine translation system for translating English lecture text to Indian languages is used. These Machine Translation systems dynamically translate or do not translate specific domain terms/ expressions based on Domain Term Identifier indication. It uses symbolic placeholders around domain terms as an indicator to the machine translation system to treat them specially and perform one of the actions of translate, transliterate or keep as it is (do not translate). For English to Indian Language Machine Translation, we first translate from English to Hindi and Telugu by ensuring that the semantic information of the source text is preserved. Hindi and Telugu were chosen since they belong to two of the prominent language families in India, namely Aryan and Dravidian, and both have considerable representation in terms of language resources compared to other Indian Languages. Most Aryan languages share similar linguistic features, for example, word order, vocabulary, aspiration, schwa deletion, etc., even though they do not share a script. Similarly, Dravidian languages are agglutinative and also share properties. Therefore to leverage this language typology, we use machine translation systems from Hindi to other considered Indo-Aryan languages and Telugu to other considered Dravidian languages for better automatic translation.

\section{Punctuation generation}
\label{sec:punctuation}
Punctuation generation is essential to provide appropriate pauses in the synthesised output so that the reproduced speech sounds natural.  There are two parts to punctuation generation, namely, a) language specific rhythm and b) speaker specific rhythm.
\subsection {Language specific rhythm}
\label{subsec:langrhythm}
An in-house Spoken token sequence Identifier that indicates how many tokens of a translation can go together in a spoken utterance without losing language specific rhythm is used. This is a specialised module on top of a language specific shallow parser (\cite{ilsl}).
As the objective is to ensure that the rhythm is performed according to the rules of a language, language specific chunkers are used. It is observed that the rhythm is more or less uniform in intra-language family. 
 
The punctuation markers provide a clue to the speech synthesis system but are still inadequate from the perspective of the rhythm of the output audio. $<audio, text>$ pairs of male and female speakers are analysed for Hindi and Tamil. Pauses in the audio and the corresponding word morpheme tags are analysed. Examples of morpheme tags for Hindi and Tamil are given in Table \ref{tab:tags}. Similar analyses are performed for other languages. The tags are used to process the text to include language specific rhythm.

\begin{table}[ht]
\centering
  \caption{A list of morpheme tags for Hindi and Tamil}
  \vspace{0.2cm}
  \label{tab:tags}
  \includegraphics[height=3cm, width=8.5cm]{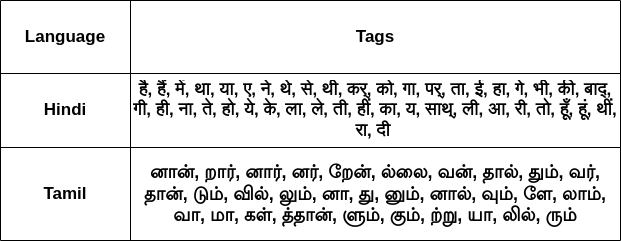}
\end{table}

\subsection {Speaker specific rhythm}
\label{subsec:spkrrhythm}
The rhythm obtained in Section \ref{subsec:langrhythm} is still inadequate for ensuring that the lecture follows the rhythm of the original speaker.  The speaker will have his/her own mannerisms in terms of pauses, based on the material that is being covered in the lecture.  The pauses in the original audio are analysed, and pauses are introduced in the target audio at proportionate locations.  The inclusion of different rhythms makes the target audio 1.2-1.5 times longer in duration.
The next step in dubbing is to sync the video and audio so that the silences in speech and the absence of lip movement in the original video are matched.

A comparison of the manual effort required was performed on 400 SWAYAM lectures (\url{swayam.gov.in}).   ASR and MT technologies reduce human effort by almost 75\%. 

\section{Text to speech synthesis in the target language and Isochronous lip syncing}
\label{sec:lipsync}
The movement of the lips in the source video is more or less based on the syllable-rate of the source speaker\cite{Park2016}.  Clearly, the source speaker's rhythm and target language rhythm must be honoured when regenerating the video. 

Target language rhythm is used to (at the SRT level) synthesise speech in the target language.   The pauses in the original audio account for speaker rhythm.   The longest silence in the original audio is first located.  The target audio is analysed so that the duration before the silence and after the silence is proportional to that of the original audio.  Syllable-based segmentation \cite{aswin2014} is used to get word boundaries.   Silence is introduced in the target audio such that the language rhythm (Section \ref{subsec:langrhythm}) is not violated.  Further, coarticulated words are also left undisturbed.  
The duration of the silence is prorated based on the duration of the target audio for every SRT. 

The video and audio durations are significantly mismatched.   Since these are technical lectures, clarity in the audio is quintessential.  Further, either the lecturer does a lot of board work or uses view graphs.    Therefore, we interpolate the video (also timestamped at the original SRT) to match the audio duration of the target language rather than the audio to the video duration.  Since the matching is performed at the syllable-level, and syllable-rate and viseme\cite{Park2016} rates are more or less equivalent, the quality of the video generated is more or less natural.
Since the word order is similar intra-language family with perhaps some variability in audio duration, the same algorithm is used for other language videos.

\section{Experiments}
\label{sec:expts}
Similar to the work reported in \cite{DBLP:journals/corr/abs-2011-03530}, we evaluate the regenerated lectures by comparing the same with the original video.  The original and target videos are between 1-2 mins long, where a small topic is covered.    We consider two types of videos: a) a lecturer with board work, b) without a lecturer and a presentation with some gestures.
The videos are dubbed in two Indian languages, namely Hindi and Tamil (Table \ref{tab:results})\footnote{\tiny{A link to the dubbed videos are available at:\\ \url{https://drive.google.com/drive/folders/1rleqYm2StSAN-xyDwwpoydbgfecPQl9G}}}.  The systems were evaluated by 17 native speakers of Hindi and 11 native speakers of Tamil.  In the video without the lecturer, it was ensured that some gestures (writing on the slide) were present.

\begin{table}[ht!]
\caption{Evaluation of dubbed videos in Hindi and Tamil}
    \label{tab:results}
    \centering
    \vspace{0.2cm}
    \begin{tabular}{|c|c|c|}
    \hline
         &  DMOS & Lip Sync\\
         \hline
        Tamil &  4.16  & 4.09 \\
        \hline
        Hindi & 4.04  & 3.51 \\
        \hline
    \end{tabular}
\end{table}
\section{Findings and Conclusions}
\label{sec:conclusion}
Informal evaluations  (Table \ref{tab:results}) suggest that the videos do give an audio-visual experience in the native tongue.  The prosody and timbre of the original speaker in terms of intonation, emphasis, and mannerisms are still missing.  Conversational speech voice conversion is still nascent \cite{bhagyashree2021}.  Prosodic modification based on domain terms may be explored to improve the quality of the videos generated.   About 70 videos have been generated in different Indian languages
\footnote{\tiny {Links to videos generated:\\ \url{https://drive.google.com/drive/folders/1hByrzH_lSceDaOQ71pWLqj6WevxaOxTz}}}.
\bibliographystyle{IEEEbib}
\bibliography{bib/refs}

\end{document}